\documentclass[10pt]{article}
\usepackage{hyperref}
\pdfoutput=1

\begin{document}

\title{The structure of a turbulent boundary layer\\ studied by numerical simulation}
\author{Philipp Schlatter, Mattias Chevalier,\\ Milo\v{s} Ilak and Dan S. Henningson \\
\\\vspace{0pt} Linn\'e FLOW Centre \\ and \\ Swedish e-Science Research Centre (SeRC) \\ KTH Mechanics, SE-100 44 Stockholm, Sweden}
\date{}

\maketitle

\begin{abstract}
Time-dependent visualisations of large-scale direct and large-eddy simulations (DNS and LES) of a turbulent boundary layer reaching up to $Re_\theta=4300$ are presented. The focus of the present fluid dynamics video is on analysing the coherent vortical structures in the boundary layer: It is clearly shown that hairpin vortices are indeed present in the simulation data, however they are characteristic remainders of the laminar-turbulent transition at lower Reynolds numbers. At higher $Re$ (say $Re_\theta>2000$), these structures are no longer seen as being dominant; the coherence is clearly lost, both in the near-wall region as well as in the outer layer of the boundary layer. Note, however, that large-scale streaks in the streamwise velocity, which have their peak energy at about half the boundary-layer thickness, are unambiguously observed. 

\vspace{0.5cm}
\noindent
Simulation data: \href{http://www.mech.kth.se/~pschlatt/DATA}{www.mech.kth.se/\~{}pschlatt/DATA}

\end{abstract}

\paragraph{Introduction} Turbulent boundary layers constitute one of the basic building blocks for understanding turbulence, particularly relevant for industrial applications. Although the geometry in technical but also geophysical applications is complicated and usually features curved surfaces, the flow case of a canonical boundary layer developing on a flat surface has emerged as an important setup for studying wall turbulence, both via experimental and numerical studies. However, only recently spatially developing turbulent boundary layers have become accessible via large-eddy simulations (LES) and even fully-resolved direct numerical simulations (DNS). The difficulties of such setups are mainly related to the specification of proper inflow conditions, the triggering of turbulence and a careful control of the free-stream pressure gradient. In addition, also the numerical cost of such spatial simulations is high due to the long, wide and high domains necessary for the full development of all relevant turbulent scales.

\paragraph{Simulation setup} In this contribution we consider a canonical turbulent boundary layer under zero pressure gradient. More specifically, visualisations of the data sets obtained via large-scale LES \cite{schlatter_li_brethouwer_johansson_henningson_2010} and DNS \cite{schlatter_orlu_2010a,schlatter_orlu_li_brethouwer_fransson_johansson_alfredsson_henningson_2009} are shown, pertaining to a spatially growing boundary layer. The inflow is a laminar Blasius boundary layer, in which laminar-turbulent transition is triggered by a random volume force shortly downstream of the inflow. This so-called trip force,  similar to a tripping strip in an experiment \cite{schlatter_orlu_li_brethouwer_fransson_johansson_alfredsson_henningson_2009}, is located at a low Reynolds number to allow the flow to develop over a long distance. Thus, the simulation covers a long, wide and high domain starting at $Re_\theta=180$ extending up to the (numerically high) value of $Re_\theta=4300$, based on the momentum thickness $\theta$ and free-stream velocity $U_\infty$. Fully developed turbulent flow is obtained starting from $Re_\theta\approx 500$.   The chosen numerical resolution for the fully spectral numerical method \cite{chevalier_schlatter_lundbladh_henningson_2007} is high enough to resolve the relevant flow structures; in the wall-parallel directions $\Delta x^+=9$ and $\Delta z^+=4$ is achieved. For the DNS, the whole simulation domain requires a total of $7.5\cdot10^9$ grid points in physical space (8192$\times$513$\times$768 spectral modes), and was thus run on a large parallel computer with 4096 processors. The well-resolved LES, using the ADM-RT model \cite{schlatter_stolz_kleiser_2004} could be run by relaxing the resolution by a factor two in each direction.

\paragraph{Flow characteristics} Turbulence statistics obtained in the boundary layer such as mean profiles, fluctuations, two-point correlations, \textit{etc.}, of the flow computed by both DNS and LES are in good agreement with other simulations and experimental studies; see the comparisons presented in Refs.\ \cite{schlatter_li_brethouwer_johansson_henningson_2010,schlatter_orlu_2010a,schlatter_orlu_li_brethouwer_fransson_johansson_alfredsson_henningson_2009}.
When it comes to spectral information recorded in the boundary layer, footprints of two very distinct spatial (and temporal) scales are detected, \textit{i.e.}, the well-known turbulent streaks close to the wall scaling in inner (viscous) units, and long and wide structures scaling in outer units. These large-scale structures persist throughout the boundary layer from the free-stream down to the wall and act as a modulation of the near-wall streaks. It is thus clear that the interplay of these two structures, scaling in a different way, is strongly dependent on the Reynolds number, \textit{i.e.}, the downstream distance.

\paragraph{Visualisation of coherent vortical structures}

However, the characteristics of these large-scale structures, and their relation to actual coherent structures present in the flow are not entirely clear, in particular for higher Reynolds numbers. Recent studies, summarised by Adrian \cite{adrian_2007} and also observed in a  DNS of low-$Re$ boundary layers \cite{wu_moin_2009} suggest a dominance of hairpin-shaped vortices of various sizes throughout the boundary layer. Hairpins are well-known structures in transitional flows, \textit{e.g.}, appearing as the results of shear-layer roll-up. It is now interesting to see whether hairpin-like structures persist in fully turbulent flow (either in the outer region or close to the wall), or whether they are mainly restricted to low-$Re$ and/or transitional flows. Thus, we would like to study the structural characterstics of a turbulent boundary layer as a function of Reynolds number, starting from the transitional phase up to fully turbulent flow at higher $Re$ than available in previous studies.

The videos in this contribution demonstrate the vortical structures arising in the boundary layer as function of the Reynolds number, $Re_\theta$, \emph{i.e.}, the downstream distance. The various parts of the video each highlight the different development stages of the boundary layer. In all frames isocontours of negative $\lambda_2$ \cite{jeong_hussain_1995} identifying vortical structures are coloured by the wall distance.  

Initially, laminar-turbulent transition is induced by trip forcing in a similar fashion as in experiments, as described above. The subsequent breakdown to turbulent flow is characterised by the appearance of velocity streaks and unambiguous hairpin vortices, which are seen to dominate the whole span of the flow, see \textit{e.g.}, Ref.\ \cite{adrian_2007}. The hairpin vortices increase in number, and individual distinctive heads of such vortices are clearly visible for some distance downstream at $Re_\theta\approx800$. This feature of low-$Re$ turbulent flow is also put forward in Ref.\ \cite{wu_moin_2009}, there denoted as ``forest of hairpins''. We can thus confirm that at least at low-$Re$ hairpins are indeed the dominant structure in a turbulent boundary layer. However, as the Reynolds number is further increased above about 1300, the scale separation between inner and outer units is getting larger, and the flow is less and less dominated by these transitional flow structures. At $Re_\theta=2500$ only isolated instances of arches belonging to hairpin vortices can still be observed riding on top of the emerging outer-layer streaky structures. But the dominance of hairpin-like structures is clearly lower than in the previous figures closer to transition.  This effect is even increased by considering the highest present Reynolds number, $Re_\theta=4300$. Then, individual hairpins or arches cannot be seen any longer. The boundary layer is now truly turbulent, and the outer layer is dominated by large-scale streaky organisation  of the turbulent vortices. Ongoing analysis of the flow structures close to the wall (\textit{i.e.}, the near-wall cycle) reveals charactersitic oscillations of the near-wall streak, mainly in a sinuous manner leading to a staggered appearance of the vortex cores \cite{schoppa_hussain_2002}.

\paragraph{Acknowledgements}
Computer time was provided by the Swedish National Infrastructure for Computing (SNIC) with a generous grant by the Knut and Alice Wallenberg Foundation (KAW). We would also like to acknowledge fruitful discussions with Qiang Li, Geert Brethouwer, Henrik Alfredsson and Arne Johansson.

\end{document}